\documentclass[reprint,preprintnumbers,amsmath,aip,amssymb]{revtex4-1}
\usepackage {amssymb,amsmath}
\usepackage {graphicx}
\usepackage{subfigure}
\usepackage[usenames]{color}
\def \be {\begin{equation}}
\def \ee {\end{equation}}
\def \ben {\begin{eqnarray}}
\def \een {\end{eqnarray}}

\usepackage{multirow}
\usepackage{ctable}

\begin {document}       
\bibliographystyle{prsty}           
  \title{Excitons: Energetics and spatio-temporal dynamics}

 \author{Seogjoo J. Jang}
 \email{seogjoo.jang@qc.cuny.edu}
 \affiliation{Department of Chemistry and Biochemistry, Queens College, City University of New York, 65-30 Kissena Boulevard, Queens, New York 11367\footnote{Primary Affiliation and Mailing Address} \& PhD programs in Chemistry and Physics, and Initiative for the Theoretical Sciences, Graduate Center, City University of New York, 365 Fifth Avenue, New York, NY 10016}
 
 \author{Irene Burghardt} 
 \email{burghardt@chemie.uni-frankfurt.de}
 \affiliation{Institute of Physical and Theoretical Chemistry, Goethe University Frankfurt, Germany}
 
 \author{Chao-Ping Hsu}
 \email{cherri@sinica.edu.tw}
 \affiliation{Institute of Chemistry, Academia Sinica, Taipei, 115, Taiwan \& Physics Division, National Center for Theoretical Sciences, Taipei, 106, Taiwan}

\author{Christopher J. Bardeen}
\email{christob@ucr.edu}
\affiliation{Department of Chemistry, University of California, Riverside, Riverside CA 92521, USA}

  \date{Published in {\it the Journal of Chemical Physics}, {\bf 155}, 200401 (2021)}

\begin{abstract}
The concept of an exciton as a
quasiparticle that represents collective excited states was originally
adapted from solid-state physics and has been successfully applied to
molecular aggregates by relying on the well-established limits of the Wannier
exciton and the Frenkel exciton.  However, the study of excitons in more complex chemical
systems and solid materials over the past two decades has  made it clear that
simple concepts based on Wannier or Frenkel excitons are not sufficient to
describe detailed excitonic behavior, especially in nano-structured solid
materials, multichromophoric macromolecules, and complex molecular aggregates.
In addition, important effects such as vibronic coupling, the influence of
charge-transfer (CT) components, spin-state interconversion, and electronic
correlation, which had long been studied but not fully understood, have turned out to play a central role in many systems.  This has motivated new experimental approaches and theoretical studies of increasing  sophistication. This article provides an overview of works addressing these issues that were published for A Special Topic of the Journal of Chemical Physics on ``Excitons: Energetics and spatio-temporal dynamics" and discusses their implications.  
\end{abstract}

\maketitle

\section{Introduction}
In a gas or dilute solution, excited electronic states are limited by the
spatial extent of an individual atom or molecule. But when multiple molecules
come together in a condensed phase environment, the boundaries and nature of
these quantum states become less well defined. The concept of an exciton as a
quasiparticle that represents these collective excited states was originally
adapted from solid-state physics and has been successfully applied to
molecular aggregates by relying on the well-established limits of the Wannier
exciton (electron-hole pair defined within band theory, with the electron-hole distance exceeding those of unit cells in a periodic system) and the Frenkel exciton (coherent
superposition of individual molecular excited states, with the electron-hole pair constituting each excitation confined within each molecule and thus within unit cells for the case of crystalline molecular aggregates).

Over the past two decades, the study of excitons in more complex chemical
systems and solid materials has attracted increasing attention because of
their importance in technologies like solar energy conversion, displays, and
sensing, just to name a few. Through these studies, it has become clear that
simple concepts based on Wannier or Frenkel excitons are not sufficient to
describe detailed excitonic behavior, especially in nano-structured solid
materials, multichromophoric macromolecules, and complex molecular aggregates.
In addition, important effects such as vibronic coupling, the influence of
charge-transfer (CT) components, spin-state interconversion, and electronic
correlation, which had long been studied but not fully understood, have turned out to play a central role in many systems.  This has motivated new experimental approaches and theoretical studies of increasing
sophistication.

Despite recent progress, there remain significant conceptual gaps
between experiment and theory, making it difficult to reach a clear consensus
concerning the detailed characteristics of excitons, in particular, their
energetics and spatio-temporal dynamics. The scope of this Special Topic was
developed with such gaps and difficulties in mind, in order to offer a timely
update and perspective on these problems. Taken together, the forty one
contributions of this Special Topic spotlight recent advances and
long-standing challenges in the spectroscopy, theory, and computation of
excitons. All of these contributions illustrate a growing level of detail and
complexity that naturally fosters the interaction between spectroscopy and
theory.  Many articles in this Special Topic would also be relevant to other
recent JCP Special Topics such as ``Singlet fission," ``2D materials," ``Up- and down-conversion in molecules and materials," and
``Polariton chemistry," which reflects the interdisciplinary and multifaceted nature of exciton research.

\section{Summary of Areas Covered}
In the following, we provide a brief overview of the main focus areas
addressed by the contributions of this Special Topic. Even though most
subtopics place emphasis on either theory/computation or spectroscopy, the
connection between these areas is a key aspect that is encountered throughout.
Indeed, the interdisciplinary nature of this field is highlighted by the fact
that the impact of a single paper can be felt across multiple areas. For
example, the introduction of a new spectroscopy technique can be accompanied
by advances in the modeling of exciton spectroscopy.

\subsection{Experimental Tools for Excitons}
Excitons are transient entities characterized by multiple time scales.  Both time and frequency domain spectroscopic methods can be used to unravel their dynamics.  Multidimensional approaches often combine time and frequency domain data to probe more detailed features of the electronic states in the optical and near-infrared spectral regions. In this collection, there are several examples of how transient absorption and time-resolved fluorescence methods can be creatively applied to learn new information about complex systems.

van Stokkum {\it et al.}\cite{vanStokkum-JCP153} use femtosecond transient absorption measurements to probe vibrational coherence and ultrafast solvation of excitonic state in covalent perylenediimide assemblies.  Powers-Riggs {\it et al.}\cite{Powers-Riggs-JCP153} also use a combination of transient absorption and fluorescence to probe symmetry breaking charge separation in noncovalent terrylenediimide assemblies. Cruz {\it et al.}\cite{Cruz-JCP153} analyze the time-resolved fluorescence signal from tetracene crystals to glean information about the spin coherence of biexciton states created by singlet fission.  Rossi {\it et al.}\cite{Rossi-JCP153} demonstrate how photoluminescence can be used to probe the origin and dynamics of dark states in perovskite quantum dots. 
These traditional nonlinear spectroscopic methods are complemented by coherent multidimensional optical spectroscopies that have been developed over the last two decades.  Collini {\it et al.}\cite{Collini-JCP154} use two-dimensional methods to map out inter- and intra-dot electronic coherences in quantum dot arrays, with an eye toward their application as quantum simulators.  Thyrhaug {\it et al.}\cite{Thyrhaug-JCP154} use two-dimensional methods to probe energy transfer and localization in a completely different system, the biological light-harvesting complex.  Maly {\it et al.}\cite{Maly-JCP153} perform a detailed 4- and 6-wave mixing study on a model  squaraine heterodimer to show that the higher order experiments can provide information about the biexciton annihilation dynamics.  Ouyang {\it et al.}\cite{Ouyang-JCP153} show that incoherent, multi-pulse nonlinear fluorescence measurements can be used to probe exciton relaxation in perovskite quantum wells.  Finally, theoretical work by Ishizaki\cite{Ishizaki-JCP153} outlines an approach for studying the spectroscopy of coupled systems by using entangled photon pairs, showing how quantum light can open up a new frontier in nonlinear spectroscopy.

A recent development in the experimental study of excitons has been the combination of high spatial resolution with the time-resolved methods described above.  This spatial resolution can be directly implemented using microscopy techniques, as described by Wittmann {\it et al.}\cite{Wittmann-JCP153} in their studies of the subdiffusive transport of singlet excitons in oligothiophene crystals.  Alternatively, Kunsel {\it et al.}\cite{Kunsel-JCP153} showed how single molecule measurements on Zn-chlorophyll aggregates can be analyzed to provide detailed information on correlated disorder on ${\rm 10\ nm}$  length scales.  It is likely that the combination of high spatial and temporal resolution methods is poised to rapidly advance our understanding of exciton dynamics in complex systems.

\subsection{Excitons from First-Principles Electronic Structure Theories}

For quantitative characterization of excitons, electronic structure calculations play an important role, using a broad range of methods adapted to molecular and solid-state systems – from density functional theory (DFT) to Green’s function approaches and excited-state quantum chemical methods. With excited state calculations, it is possible to predict many characteristics of excitations in molecular aggregates as well as inorganic or nano-carbon materials, with reasonable accuracy, as demonstrated in works by Patternson,\cite{Patterson-JCP153} Valente {\it et al.},\cite{Valente-JCP154} Li {\it et al.}\cite{Li_X-JCP153}, and Yang {\it et al.}\cite{Yang_Y-JCP153}  Given the size of the molecular systems in typical studies, time-dependent density functional theory (TD-DFT) is still the method of choice in most cases. In cases where multi-exciton states are crucial, as in singlet fission, approximate multi-reference methods like multireference configuration interaction DFT (DFT/MRCI) have been explored.\cite{Valente-JCP154}  In addition to the chosen electronic structure method, the geometry dependence of the relevant electronic states is crucial as demonstrated by Deutsch {\it et al.}\cite{Deutsch-JCP153} and Mukazhanova {\it et al.},\cite{Mukazhanova-JCP153} especially if complex excited-state relaxation and reorganization processes are involved, possibly leading to the formation of traps.

Recent first-principles studies have often revealed the existence and importance of CT components in low-lying excited states. A small CT admixture can lead to an enhancement of low electronic coupling for singlet fission, while large CT components have also been found to play an important role in the spectra of crystals, aggregates, and light-harvesting complexes in photosynthetic apparatuses.  Long-range corrected (LC) DFT is a good compromise for charge transfer characteristics, as illustrated in the tests of LC-DFTB by Darghouth {\it et al.}\cite{Darghouth-JCP154}  Other discussions on the CT components are seen in the works by Sun {\it el al.},\cite{Sun-JCP153}  Balooch Qarai {\it et al.},\cite{BaloochQarai-JCP153} and others.\cite{Li_X-JCP153,Valente-JCP154,Thyrhaug-JCP154,Klinger-JCP153} 

With electronic states obtained, T\"{o}lle {\it et al.}\cite{Tolle-JCP153} show how diabatic states can be defined and calculated, from which an excitonic model Hamiltonian can be built. The resulting {\it ab initio} exciton models\cite{Li_X-JCP153}  are a key step to connect to model Hamiltonians for exciton photophysics and transport that have been employed for decades. Ramos and Pavanello\cite{Ramos-JCP154}  show that constrained DFT  is a useful technique to generate diabatic states and to determine the electronic coupling in the diabatic basis.  
In a similar vein, another important advance concerns the multi-state treatments for obtaining a full set of matrix elements, which allows a more complete description of the exciton dynamics, with Frenkel excitons, CT excitons, and multi-excitons included in the model.\cite{Li_X-JCP153,Tolle-JCP153} Various contributions in this Special Topic collection illustrate the flexibility and quantitative performance of vibronic exciton models that can be constructed from such multi-state descriptions.\cite{BaloochQarai-JCP153,Sun-JCP153,Yang_L-JCP153,DiMaiolo-JCP153}  

\subsection{Exciton Rates and Lineshape Theories}
Transfer of localized excitons has traditionally been described by Förster rate theory, which is rooted within the standard perturbation theory framework of Fermi’s golden rule.  However, recent works highlight the importance of mechanisms\cite{Collini-JCP154,DiMaiolo-JCP153,Cainelli-JCP154} that go beyond the Förster theory and the possibility of how they act to accelerate exciton transport.  Yang and Jang\cite{Yang_L-JCP153} provide detailed modeling of such effects due to common vibrational modes within the assumption of rate behavior.  Beyond rate theories, applications of more advanced quantum dynamics approaches are necessary.\cite{DiMaiolo-JCP153,Cainelli-JCP154}   Yet, depending on the parameter regime, exciton transfer processes in many cases appear to exhibit rate behavior even when there is no apparently small parameter that justifies the standard perturbation theory result. Seibt {\it et al.}\cite{Seibt-JCP153} address this issue by comparing rates defined within various approximations with the exact rates that can be extracted from the hierarchical equations of motion (HEOM) approach. The authors highlight the importance of correctly describing nonequilibrium initial conditions and compare the HEOM rates with a cumulant approximation, providing a simple approach  accounting exclusively for the effects of fluctuations for the calculation of exciton transfer rates. 

Electronic couplings and exciton-phonon interactions, which are typically taken to be fixed in calculating exciton transfer rates, can in fact be subject to fluctuations, for example, due to exciton-exciton interactions and other electronic correlation effects. More generally, excitons in various environments are subject to multitudes of many-body potentials and perturbations that are difficult to represent by simple forms of Hamiltonians amenable to dynamics calculations. In the early days of exciton research, such effects were treated phenomenologically by introducing stochastic fluctuations with prescribed properties at the level of the exciton Hamiltonian.  While successful in describing the observed phenomena, such an approach does not necessarily provide satisfactory understanding of the source and nature of fluctuations.  Srimath Kandada {\it et al.}\cite{Kandada-JCP153} bring the stochasticity down to a more fundamental level and introduce it in the background exciton density.  Combining a field theoretic formulation and stochastic calculus, a theory of lineshape is developed that generalizes the Anderson-Kubo theory.  The new theory is capable of producing nonlinear spectral shifts and asymmetries in lineshape.  The theory can be refined further to represent a more general class of stochastic fluctuations. In a related effort, Wang {\it et al.}\cite{Wang-JCP153} develop a theory of emission power spectra in the framework of macroscopic quantum electrodynamics, including exciton-photon and exciton-phonon interactions in the presence of arbitrary media while distinguishing between coherent and incoherent regimes.

\subsection{Quantitative Modeling of Exciton Spectroscopy}
The rapid advances in experimental spectroscopic methods have been accompanied by a rapid growth in the ability of theory to analyze the results, and most of the experimental papers described in subsection A include a theoretical component to help model the data.\cite{Wittmann-JCP153,Powers-Riggs-JCP153,vanStokkum-JCP153,Thyrhaug-JCP154}   Our ability to measure and interpret excitonic spectra has advanced to the point where such spectra can be used as a diagnostic to monitor molecular assembly and conformational changes, as shown by the work of Sosa {\it et al.}\cite{Sosa-JCP153} studying the formation of pseudoisocynanine thin films.  Further, highly efficient simulation protocols for $n$th-order spectroscopic signals have been developed, exemplified by the automated Feynman diagram generator presented by Rose {\it et al.}\cite{Rose-JCP154} Nevertheless, various computational studies show that achieving quantitative agreement between first-principles modeling and the measured dynamics remains challenging due to the high sensitivity of excitonic features to small energy differences as well as shortcomings in the underlying electronic structure description and the dynamical computation scheme.  Even for simple absorption spectra of dimers of organic dye molecules in solution phase, Kumar {\it et al.}\cite{Kumar-JCP154} show that careful accounts of structural and energetic disorder and fluctuations as well as employing high level electronic structure calculation and quantum dynamics methods are necessary for accurate modeling.

Annihilation of excitons, where two excitons of similar energy interact with each other resulting in the disappearance of one, has long been recognized as playing an important role in multiexciton dynamics. On the experimental side, exciton-exciton annihilation is employed to measure diffusion coefficients. Detailed theoretical understanding and developing ways to experimentally control this process also have significant technological implications for new solar energy conversion devices and laser technology.  Süß and Engel\cite{Sus-JCP153} offer a new theoretical understanding of this process by developing detailed wave packet dynamics for a trimer system with two different initial separations of excitons and explicitly calculating 2D-electronic spectroscopy signals that are fifth order with respect to the matter-radiation interaction.  The results of these calculations suggest the possibility to detect signatures of different initial conditions and directly monitor coherence between excited states participating in the annihilation process.   

Many new concepts rely on the combination of spectroscopy and theory. For example, quantum light, with properly designed photon entanglement, can open new and less intrusive ways to probe the energetics and spatiotemporal dynamics of excitons.  The theoretical formulation for this, let alone experimental demonstration, is as yet in an early stage but promising.  For sufficiently short entanglement times, it is shown that frequency entangled photon pairs generated from a single frequency laser source can temporally resolve the state-to-state dynamics of excitons through transmission spectroscopy.\cite{Ishizaki-JCP153} The information gained this way is equivalent to heterodyne four wave mixing spectroscopy and suggests that more elaborate engineering of quantum photons may provide new information that is not accessible through classical light.

The description and spectroscopy of polaritons, resulting from strong light-matter interaction, are another important emergent field. In this context, Alvertis {\it et al.}\cite{Alvertis-JCP153} report on computational modeling of bulk exciton-polaritons formed in organic materials.  For this, first principle calculations along with electronic structure calculation are desirable, which however remains a major computational challenge.  

\subsection{Quantum Dynamics of Excitons}

Molecular excitons are quantum mechanical objects and in general require quantum approaches to describe their dynamics.  Therefore, it is no surprise that the exciton dynamics have become one of the primary fields of applications of various quantum dynamics approaches developed in chemical physics. These approaches can be classified into two kinds, one of which is rooted in system-bath theory for the electronic versus vibrational degrees of freedom, while the other approach endorses a combined quantum or quantum-classical evolution in the full electronic-plus-vibrational space. Both schemes often start with rather simple model Hamiltonians where the small number of excitonic states are considered as “system” whereas all other degrees of freedom (viewed as “bath”) can be modeled, for example, by an infinite number of harmonic oscillators defining a spectral density. This results in a general class of vibronic coupling Hamiltonians.\cite{BaloochQarai-JCP153,Yang_L-JCP153,DiMaiolo-JCP153} Today, this approach is rendered far more quantitative by the first-principles computation of spectral densities, for example, as demonstrated by Klinger {\it et al.}\cite{Klinger-JCP153} and ultimately by the connection to {\it ab initio} exciton Hamiltonians.\cite{Li_H-JCP153,Kumar-JCP154} Quantum master equations (QME) and their generalization, such as the HEOM or path integral influence functional approach, have been particularly successful in the first – {\it i.e.}, system-bath type -- approach. The second, alternative approach has made significant progress, too, {\it e.g.}, employing an advanced semiclassical dynamics method,\cite{Kumar-JCP154} relying on efficient multiconfigurational wavefunction propagation in the full system-plus-bath space using the multi-layer multiconfiguration time-dependent Hartree (ML-MCTDH) technique, or using quantum-classical schemes based on the quantum-classical Liouville equation (QCLE). 

In this collection, there are works in both of these directions. Cainelli and Tanimura\cite{Cainelli-JCP154} apply the HEOM approach to investigate the role of local vs. non-local exciton-phonon coupling, and Yan {\it et al.}\cite{Yan-jcp153} report further improvement in the HEOM approach. Di Maiolo {\it et al.}\cite{DiMaiolo-JCP153}  show the promise of the ML-MCTDH method as a full quantum propagation scheme for hundreds of electronic states and vibrational modes, which is applied to coherent intra-chain exciton migration and the computation of temperature-dependent diffusion coefficients.  Sun {\it et al.}\cite{Sun-JCP153} employ a variational, multiconfigurational Davydov approach to probe the temperature dependence of singlet fission. Mixed quantum-classical nonadiabatic quantum dynamics approaches based on the QCLE are addressed by Kim and Rhee;\cite{Kim-JCP153} these can potentially be generalized for a large class of molecular systems but rely on certain kinds of approximations that become reasonable in the semiclassical or classical limits of nuclear degrees of freedom.  Finally, Acharyya {\it et al.}\cite{Acharyya-JCP153} report application of a path integral approach for donor-bridge-acceptor (DBA) model with off-diagonal couplings.  Although this work more specifically addresses electron transfer process, the observed results are equally effective in DBA exciton transfer system. 

Following an alternative route, direct {\it ab initio} quantum dynamics simulation of exciton dynamics has become a realistic option, both for inorganic semiconductors using {\it ab initio} real-time TDDFT\cite{Yang_Y-JCP153} and also for many molecular systems in complex environments.\cite{Freixas-JCP153} This progress is due to the improvement in both accuracy and efficiency of electronic structure calculation methods along with the development of approximate nonadiabatic quantum dynamics methods with reasonable accuracy.  Freixas {\it et al.}\cite{Freixas-JCP153} demonstrate this capability and as yet challenging practical issues through direct dynamics simulations for a few representative molecular triads in solution phase, employing the non-adiabatic molecular dynamics (NAMD) software package developed by the Los Alamos group led by Tretiak and coworkers.  The authors of this work simulate various exciton dynamics by combining surface hopping methods with on-the-fly time dependent density functional or Hartree-Fock calculations. At the same time, they also clarify that the electronic structure calculation methods employed, while being able to model and explain conventional spectroscopic data with reasonable accuracy, tend to fall short of offering quantitatively reliable information on the exciton dynamics due to the sensitivity of the dynamics calculations to even small errors in electronic energies and coupling constants.  These results show that black-box application of any nonadiabatic computational package, without detailed knowledge of potential errors and misinterpretation, is not yet possible. We anticipate that a combination of direct ab initio quantum dynamics and first-principles parametrized exciton Hamiltonians could offer a viable future route towards the modeling of nonadiabatic dynamics in extended systems.

\subsection{Nanostructured Systems}

The connection between theory and experiment relies on an accurate knowledge of the structure of the system.  Single crystals\cite{Cruz-JCP153,Wittmann-JCP153}  provide a bulk system whose structure is well-defined, while molecular assemblies, both covalent\cite{vanStokkum-JCP153,Maly-JCP153} and noncovalent,\cite{Sosa-JCP153,Powers-Riggs-JCP153,Kunsel-JCP153,Thyrhaug-JCP154}  provide nanostructures whose intermolecular interactions can be modified using synthetic methods.  The growth of inorganic systems can provide well-defined structures in the form of quantum wells\cite{Ouyang-JCP153} or quantum dot arrays.\cite{Collini-JCP154,Rossi-JCP153}   A new way to create structurally well-defined exciton states is to introduce localized chemical defects into two-dimensional monolayers.   For example, tungsten atom doping in ${\rm MoSe_2}$ affects the electron-hole pair dynamics.\cite{Yang_Y-JCP153} Smiri {\it et al.}\cite{Smiri-JCP154} also provide detailed computational studies on the properties of strained nanobubbles in ${\rm MoS_2}$. 

How to best design artificial light harvesting system remains a highly active area of research both theoretically and experimentally. Here, too, highly ordered morphologies play a key role. While there have been significant number of suggestions and sometimes anecdotal evidence of modest improvement through design, significant enhancement in the ability to transport excitons from one site to another distant site - a key process for efficient solar energy conversion - has been difficult to identify.  In this context, Davidson {\it et al.}\cite{Davidson-JCP153}  provide such an example for one-dimensional nanowire architectures that is supported through extensive simulation.  Contrary to the conventional wisdom, this work demonstrates that embedding spikes of high exciton states in overall downhill energy gradient can result in significant enhancement in the efficiency of exciton transport.  The reason is that dark states created by inserting those high energy states protect the exciton against super-radiance.  This design concept is easy to understand and can be implemented experimentally.  

\section{Conclusion}
The articles in this Special Topic provide a good cross-section of current research in this rapidly advancing field.  They demonstrate remarkable advances made in all three aspects of the spectroscopy, electronic structure calculation, and dynamics/kinetics simulation of excitons to the extent that interesting and plausible analyses become possible. However, it is clear that there remain significant challenges for both theory and experiment in attaining satisfactorily quantitative understanding of excitons in complex chemical systems.  From a fundamental standpoint, such an understanding would clarify the spatial extent of quantum states and how they interact with both light and the environment.  From a technological standpoint, such understanding will be crucial for the development of materials and characterization methods in fields ranging from solar energy conversion to quantum information science.  Integration of experimental and theoretical studies to cross-examine and validate key assumptions and interpretations is necessary to accomplish this goal. We hope this collection of papers will help advance these efforts in the physical chemistry and chemical physics community.

\acknowledgments

We thank all the authors who contributed; Tim Lian for suggesting and encouraging this Special Topic, Jennifer Ogilvie, David Reichman, Emily Weiss, and Xiaoyang Zhu for managing the review process, Erinn Brigham and Judith Thomas for editorial assistance.  SJJ acknowledges major support from the US Department of Energy, Office of Sciences, Office of Basic Energy Sciences (DE-SC0021413) and for additional support from the US National Science Foundation (CHE-1900179).  IB acknowledges support from the Deutsche
Forschungsgemeinschaft (DFG grant BU-1032-2).  CPH acknowledges supports from Academia Sinica Investigator Award (AS-IA-106-M01) and the Ministry of Science and Technology of Taiwan (projects 109-2113-M-001-022-MY4 and MOST 110-2123-M-001-005).  CJB acknowledges support from the US National Science Foundation (CHE-1800187).

\end{document}